\documentclass[aps,pre]{revtex4}
\usepackage{graphicx}
\begin{document}

\title{Exact relations between damage spreading and thermodynamic functions
for the N-color Ashkin-Teller model}
\author{A. S. Anjos$^{1}$}
\author{I.S. Queiroz$^{2}$}
\author{A.M. Mariz$^{1}$}
\author{F.A. da Costa$^{1,}$}\altaffiliation{Corresponding author: fcosta@dfte.ufrn.br}
\affiliation{$^{1}$ Departamento de F\'{\i}sica Te\'orica e Experimental, Universidade Federal do Rio Grande do Norte \\ Caixa Posta 1641, 59072-970 Natal--Rio Grande do Norte, Brazil, \\
$^{2}$ Departamento de Ci\^encias Ambientais
Universidade Federal Rural do Semi-\'Arido - UFERSA, \\ BR 110 - Km 47, Bairro Presidente Costa e Silva,
CEP 59.625-900 Mossor\'o--Rio Grande do Norte,Brazil}

\begin{abstract}
Some exact results are derived relating quantities computable by the so-called damage spreading method and
thermodynamic functions for the N-color Ashkin-Teller model. The results are valid for any ergodic dynamics. Since we restrict our analysis to the ferromagnetic case the results are also valid for any translational invariant lattice. The derived relations should be used in order to determine numerically the N-color Ashkin-Teller critical exponents with better accuracy and less computational efforts than standard Monte Carlo simulations.
\end{abstract}

\maketitle

Since its introduction, the damage spreading (DS) method has become a powerful tool in the study of phase transitions and critical phenomena \cite{stan87,der87}. Basically it consists in following the time evolution of  two initially identical copies of a given system, both subject to the same dynamical rules and to the same thermal noise, after the introduction of a small perturbation (called “damage”) in one of them at a given time. This method presents some advantages on the traditional Monte Carlo (MC) method where the time evolution of a single copy of the system is investigated \cite{lb2000}. For instance, in the DS method the fluctuations are substantially reduced as compared to the conventional MC method. A recent review on the DS method can be found in \cite{puzzo08}.

In statistical mechanics, the most important question when performing a simulation is how to obtain the equilibrium thermal properties of the system under investigation. Thus it is very important to establish exact relations relating measurable quantities obtained in a DS simulation to the thermodynamical properties of the system. Such relations were first obtained for the ferromagnetic Ising model in a square lattice and the numerical results showed a prominent reduction of fluctuations and finite-size effects \cite{carj89}. In the wake of those findings the exact relations between damage spreading computable quantities such as magnetization and spin-spin correlations functions and thermal equilibrium quantities has been obtained for several systems like, for instance, the Potts \cite{mariz90}, Ashkin-Teller \cite{mariz90}, discrete N-vector \cite{mariz93}, $(N_{\alpha}, N_{\beta})$ \cite{mariz98}, and spin-1 \cite{anjos08}  models. These relations are valid for any translationally invariant system which evolves in time under any ergodic dynamics. Recently, such relations were applied to numerically investigate the Potts \cite{anjos06}, two-color Ashin-Teller \cite{anjos07} and a spin-1 \cite{anjos08} models. In all these cases the computed critical exponents were in aggreement to the exact known results to within four decimal places, revealing the accuracy of the DS method.

The Ashkin-Teller (AT) model was introduced to investigate cooperative phenomena in quaternary alloys \cite{at43}. It was shown to be equivalent to a two-layer Ising model with a four-spin coupling between the two layers \cite{fan72}. In two dimension the AT model can be mapped onto a staggered eight-vertex model at the critical point and shows non-universal critical behavior along a self-dual line where the exponents vary continously \cite{baxter82}.  In general, despite its simplicy the two-color AT model displays a wide variety of critical and multicritical phenomena already in two dimension, as has been shown in studies based on duality arguments \cite{wulin74}, real-space renormalization group \cite{domany79, mariz85, bezerra01}, mean-field renormalization-group approachs \cite{pla86, pmco89, pla99}, finite-size scaling \cite{kamieniarz97, pawlicki97, badehdah00}, and conventional Monte Carlo simulations \cite{kamieniarz97, ditzian80, chahine89, wiseman93,  bakechi99}.

The N-color Ashkin-Teller (N-AT) model consists of $N$ Ising models coupled pairwise through a four-spin interaction \cite{grest81}. When $N=2$ it becomes the usual AT model discussed above
(also known as two-color AT model). It was argued that for $N > 2$ the model has a first-order transition as long as the four-spin coupling is ferromagnetic \cite{grest81}. In \cite{grest81} the phase diagram was investigated by mean-field analysis and Monte Carlo simulation for $N=3$ and by several other tecniques for $N > 2$  \cite{fradkin84, bray85, kardar85, shankar85, goldschmidt86, martins88, defelicio03, piolho08}. Recently there is a renewed interest in this model mainly due to the richness of its phase diagram and its relation to other systems \cite{calabrese02, calabrese04, papa07, goswami08}. However, it still lacks a lot of work in order to fully understand the main features of the N-AT for $N > 2$ in any dimension. We believe that the DS simulations are quite suitable to give us more information about this model.

The purpose of the present note is to determine exact relations envolving thermodynamic quantities and some specific combinations of damages for the N-color Ashkin-Teller model. The model is defined by the Hamiltonian

\begin{equation}
\mathcal{H} = - J_2 \sum_{\langle i j \rangle} \sum_{a=1} ^{N} \sigma_i^a \sigma_j^a - \frac{1}{2}J_4 \sum_{\langle i j \rangle} \sum_{a \ne b}^{N} \sigma_i^a \sigma_i^b \sigma_j^a \sigma_j^b ,
\end{equation}

\noindent
where $\sigma_i^a = \pm 1$,  $a$ and $b$ labels different Ising spins (distinguished by their ``color'') at a site $i$ and $\langle i j \rangle$ indicates that the sum is performed over all distinct pairs of sites on a given lattice. For $N=1$ we recover the Ising model, whereas for $N=2$ we have the usual Ashkin-Teller model.  In what follows, our treatment is valid for any value of $N$, and for $(J_2, J_4)$ in the ferromagnetic range in order to take into account the translational invariance of the lattice.

According to the usual mean-field procedure we can write a single-site effective Hamiltonian

\begin{equation}
\beta \mathcal{H}_{eff} = \sum_{a=1}^{N} h_{i}^a \sigma_i^a + \frac{1}{2} \sum_{a \ne b}^N h_i^{ab} \sigma_i^a \sigma_i^b ,
\end{equation}

\noindent
where $h_i^a = K_2 \sum_{j \ne i} \langle \sigma_j^a \rangle$ and $h_i^{ab} = K_4 \sum_{j \ne i} \langle \sigma_j^a \sigma_j^b \rangle$. In the above equation  $\langle \cdots \rangle$ denotes thermal average and where we have introduced the new variables $K_n = J_n/k_{B}T$ ($n= 2, 4$). The structure of the effective fields acting on $\sigma_i^a$ and $\sigma_i^a \sigma_i^b$ implies that the order parameters are given by

\begin{equation} \label{ma}
m^a = \langle \sigma_i^a \rangle , \quad a = 1, \cdots, N,
\end{equation}

\noindent
and
\begin{equation} \label{Mab}
M^{ab} = \langle \sigma_i^a \sigma_i^b \rangle , \quad a \ne b,
\end{equation}

\noindent
while the correlation functions are expressed by

\begin{equation} \label{Ga}
\Gamma_{ij}^a = \langle \sigma_i^a \sigma_j^a \rangle - \langle \sigma_i^a \rangle  \langle \sigma_j^a \rangle, \quad a = 1, \cdots, N,
\end{equation}

\noindent
and

\begin{equation} \label{Gab}
\Gamma_{ij}^{ab} = \langle \sigma_i^a \sigma_i^b \sigma_j^a \sigma_j^b \rangle - \langle \sigma_i^a \sigma_i^b \rangle  \langle \sigma_j^a \sigma_j^b \rangle, \quad a \ne b.
\end{equation}

To implement the damage spreading technique it is convenient to introduce the following binary variables

\begin{eqnarray}
\Pi_i^a & = & \frac{1}{2} (1 + \sigma_i^a), \label{Pa} \\
\Pi_i^{ab} & = & \frac{1}{2} (1 + \sigma_i^a \sigma_i^b), \quad a \ne b . \label{Pab}
\end{eqnarray}

Let us consider two configurations (referred to as $A$ and $B$) on a regular, translational invariant, lattice evolving in time according to the same (and ergodic) dynamics such as Metropolis, heat-bath or Glauber \cite{lb2000}. We define four different types of damage combinations:

\begin{eqnarray}
&(1)&  \Pi_i^a (A) = 1 \quad \mathrm{and} \quad \Pi_i^a (B) = 0  , \\
&(2)&  \Pi_i^a (A) = 0 \quad \mathrm{and} \quad \Pi_i^a (B) = 1  , \\
&(3)&  \Pi_i^{ab} (A) = 1 \quad \mathrm{and} \quad \Pi_i^{ab} (B) = 0,  ~~ a \ne b , \\
&(4)&  \Pi_i^{ab} (A) = 0 \quad \mathrm{and} \quad \Pi_i^{ab} (B) = 1,  ~~ a \ne b .
\end{eqnarray}

It is known that after a long time the system eventually reach the thermal equilibrium. In this regime, the above-defined damages occur with probabilities given, respectively, by:

\begin{eqnarray}
p_1 & = & \displaystyle{ \langle \Pi_i^a(A) \left[ 1 - \Pi_i^a(B)\right] \rangle_t } ~, \\
p_2 & = & \displaystyle{ \langle  \left[ 1 - \Pi_i^a(A)\right] \Pi_i^a(B) \rangle_t } ~, \\
p_3 & = & \displaystyle{ \langle \Pi_i^{ab}(A) \left[ 1 - \Pi_i^{ab}(B)\right] \rangle_t } ~, \\
p_4 & = & \displaystyle{ \langle  \left[ 1 - \Pi_i^{ab}(B)\right] \Pi_i^{ab}(B) \rangle_t } ~,
\end{eqnarray}

\noindent
where $ \langle \cdots \rangle_t $ means time average over the trajectory followed by the copies of the system in their phase space. In what follows it is convenient to introduce differences between such probabilities:

\begin{equation} \label{F}
F = p_1 - p_2 = \langle \Pi_i^a(A) \rangle_t -  \langle \Pi_i^a(B) \rangle_t ~, ~~ a = 1, \cdots, N,
\end{equation}

\noindent
and

\begin{equation} \label{G}
G = p_3 - p_4 = \langle \Pi_i^{ab}(A) \rangle_t -  \langle \Pi_i^{ab}(B) \rangle_t ~, ~~ a \ne b .
\end{equation}

The next step in our analysis consists in imposing some constraints in the temporal evolution. In the present case it turns out that there are four distinct such possibilities:

\begin{description}
\item[($e_1$)] copy $A$ evolves without any constraint, while copy $B$ is restricted to the constraint that for an arbitrarily fixed site, say $i = 0$, $\Pi_0^a (B) = 0 $, for $ a = 1, \cdots, N $.
\item[($e_2$)] copy $A$ evolves with $\Pi_0^a(A) = 1$ and copy $B$ with $\Pi_0^a(B) = 0 $, for $~ a = 1, \cdots, N $.
\item[($e_3$)] copy $A$ evolves without any constraint, while copy $B$ is restricted to with $\Pi_0^{ab}(B) = 0, ~ a \ne b $.
\item[($e_4$)] copy $A$ is subjected to $\Pi_0^{ab}(A) = 1$, while copy $B$ is restricted to $\Pi_0^{ab}(B) = 0, ~~ a \ne b $.
\end{description}

Ergodicity implies, for the evolution $(e_1)$, that

\begin{equation} \label{erg1}
\langle \Pi_i^a(A) \rangle_t = \langle \Pi_i^a \rangle
\end{equation}

\noindent
and, with the help of conditional probability,

\begin{equation} \label{erg2}
\langle \Pi_i^a(B) \rangle_t = \frac{\langle \Pi_i^a \left(1 - \Pi_0^a  \right) \rangle}{1 - \langle \Pi_0^a \rangle} ~.
\end{equation}

Eqns (\ref{erg1},\ref{erg2}) relates, in a definitive way, the dynamical and thermal averages represented, respectively, by $\langle \cdots \rangle_t$ and $\langle \cdots \rangle$. Substitution of those equations into (\ref{F}) gives us

\begin{equation}
F(e_1) = \frac{\langle \Pi_i^a \Pi_0^a \rangle - \langle \Pi_i^a \rangle \langle \Pi_0^a\rangle ~}{1 -\langle \Pi_0^a\rangle }
\end{equation}

\noindent
which, with the help of (\ref{ma}), (\ref{Ga}) and  (\ref{Pa}), can be expressed as

\begin{equation} \label{fe1}
F(e_1) = \frac{\Gamma_{0i}^{a}}{2(1 - m^a)} ~.
\end{equation}

For the evolution $(e_2)$ we have, in a similar way,

\begin{equation} \label{erg3}
\langle \Pi_i^a (A) \rangle_t = \frac{\langle \Pi_0^a \Pi_i^a \rangle ~}{\langle \Pi_0^a \rangle} ,
\end{equation}

\noindent
and

\begin{equation} \label{erg4}
\langle \Pi_i^a (B) \rangle_t = \frac{\langle \Pi_i^a \left(1 -
\Pi_i^a\right) \rangle ~}{1 - \langle \Pi_0^a \rangle} ,
\end{equation}

\noindent
from which follows

\begin{equation} \label{fe2}
F(e_2) = \frac{\Gamma_{0i}^a ~}{1 - (m^a)^2} .
\end{equation}

\noindent
In order to write the final Eqs. (\ref{fe1}) and (\ref{fe2}) we have used the fact that $m^a = m_0^a $ as a consequence of the lattice translation invariance. The results expressed by these equations means that at first we compute numerically $F(e_1)$ and $F(e_2)$ (quantities related to DS). Then, we may determine the magnetization and the two-point correlation functions (thermal equilibrium quantities) as

\begin{equation}
m^a = 2 \frac{F(e_1)}{F(e_2)} - 1 ,
\end{equation}

\noindent
and

\begin{equation} \label{g2a}
\Gamma_{0i}^a = 4 \frac{F(e_1)}{F(e_2)} \left( F(e_2) - F(e_1)
\right) ,
\end{equation}

\noindent
for $ a = 1, 2, \cdots, N $.

For the evolution $(e_3)$ we find

\begin{equation}
\langle \Pi_i^{ab} (A) \rangle_t = \langle \Pi_i^{ab} \rangle
\end{equation}

\noindent
and

\begin{equation}
\langle \Pi_i^{ab} (B) \rangle_t = \frac{\langle \Pi_i^{ab} (1 - \Pi_o^{ab}) \rangle ~}{1 - \langle \Pi_0^{ab} \rangle} ,
\end{equation}

\noindent
for $ a \ne b $.

Thus, from (\ref{Mab}), (\ref{Gab}) and (\ref{Pab}) we get

\begin{equation} \label{ge3}
G(e_3) = \frac{\Gamma_{0i}^{ab}}{2(1 - M^{ab})} , ~~ a \ne b .
\end{equation}

Finally, evolution $(e_4)$ implies that

\begin{equation}
\langle \Pi_i^{ab}(A) \rangle_t = \frac{\langle \Pi_i^{ab} \Pi_0^{ab} \rangle}{\langle \Pi_0^{ab} \rangle}
\end{equation}

\noindent
and

\begin{equation}
\langle \Pi_i^{ab}(B) \rangle_t = \frac{\langle \Pi_i^{ab}(1 - \Pi_0^{ab}) \rangle}{1 - \langle \Pi_0^{ab} \rangle} ,
\end{equation}

\noindent
for $a \ne b $.

Therefore, in terms of $M^{ab}$ and $\Gamma_{0i}^{ab}$ we have

\begin{equation} \label{ge4}
G(e_4) = \frac{\Gamma_{0i}^{ab}~}{1 - (M^{ab})^2} , ~~ a \ne b .
\end{equation}

Thus, having determined $G(e_3)$ and $G(e_4)$ in a numerical simulation, we can use (\ref{ge3}) and (\ref{ge4}) to compute

\begin{equation}
M^{ab} =  2 \frac{G(e_3)}{G(e_4)} - 1
\end{equation}

\noindent
and

\begin{equation}
\Gamma_{0i}^{ab} =  4 \frac{G(e_3)}{G(e_4)} \left( G(e_4) - G(e_3) \right) .
\end{equation}

Since $F(e_1)$, $F(e_2)$, $G(e_3)$ and $G(e_4)$ are computed from selected combinations of damages, the relations expressed by (\ref{fe1} - \ref{fe2}) and (\ref{ge3} - \ref{ge4}), which are exact for any translationally invariant lattice at all temperatures, allow us to determine numerically the thermal properties of the $N$-color Ashkin-Teller model.

As an example we considered the ferromagnetic 3-color Ashkin-Teller model on a square lattice of linear size $L = 60$ and with $J_4 = 0.\!01J_2$. Periodic boundary condition were applied and we computed the two-site correlation functions

\begin{equation}
 \Gamma^{(a)}(r) = \frac{1}{4} \sum_{i(r)} \Gamma^{(a)}_{0i} \quad (a = 1, 2, 3) , 
\end{equation}

\noindent measured with respect to the central site located at $(L/2,
L/2)$. The two-site correlation functions $\Gamma^{(a)}_{0i}$ are
computed from (\ref{g2a}).  As the equilibrium was approached we
found that $ \Gamma^{(1)}(r) = \Gamma^{(2)}(r) = \Gamma^{(3)}(r) $
as expected from the permutation symmetry between the spin
variables presented by the Hamiltonian. Initially a copy denoted
by $A$ with all spin variables $\sigma^{(a)}_i(A) = 1$ is left to
evolve according to a MC run during a time $t_{eq} = 1 \times
10^4$ which is sufficient to the system get close enough to
thermal equilibrium. Then a second copy denoted by $B$ is created
by replicating copy $A$ and introducing small modifications, the
initial damage, on it. Both copy are left to evolve under the same
ergodic dynamical rules during a time $t_{eq} = 1.\!8 \times 10^5$
and subjected to specific boundary conditions corresponding to the
possibilities ($e_1$) and ($e_2$) above mentioned. In addition to
that, each simulation were performed for $M = 20$ different
samples in order to reduce inherent statistical fluctuations for a
given temperature. To locate the transition point we search for
the temperature ratios $T/T_C$ at which the functions
$\Gamma^{(a)}(r)$ show the slowest decay as a function of $r$. At
criticality we expect that, for large $r$,

\begin{equation}
\Gamma^{(a)}(r) \sim r^{-\eta} \quad (a = 1, 2, 3).                                                     \end{equation}

\begin{figure}
\begin{center}
\includegraphics[angle=-90,scale=0.5]{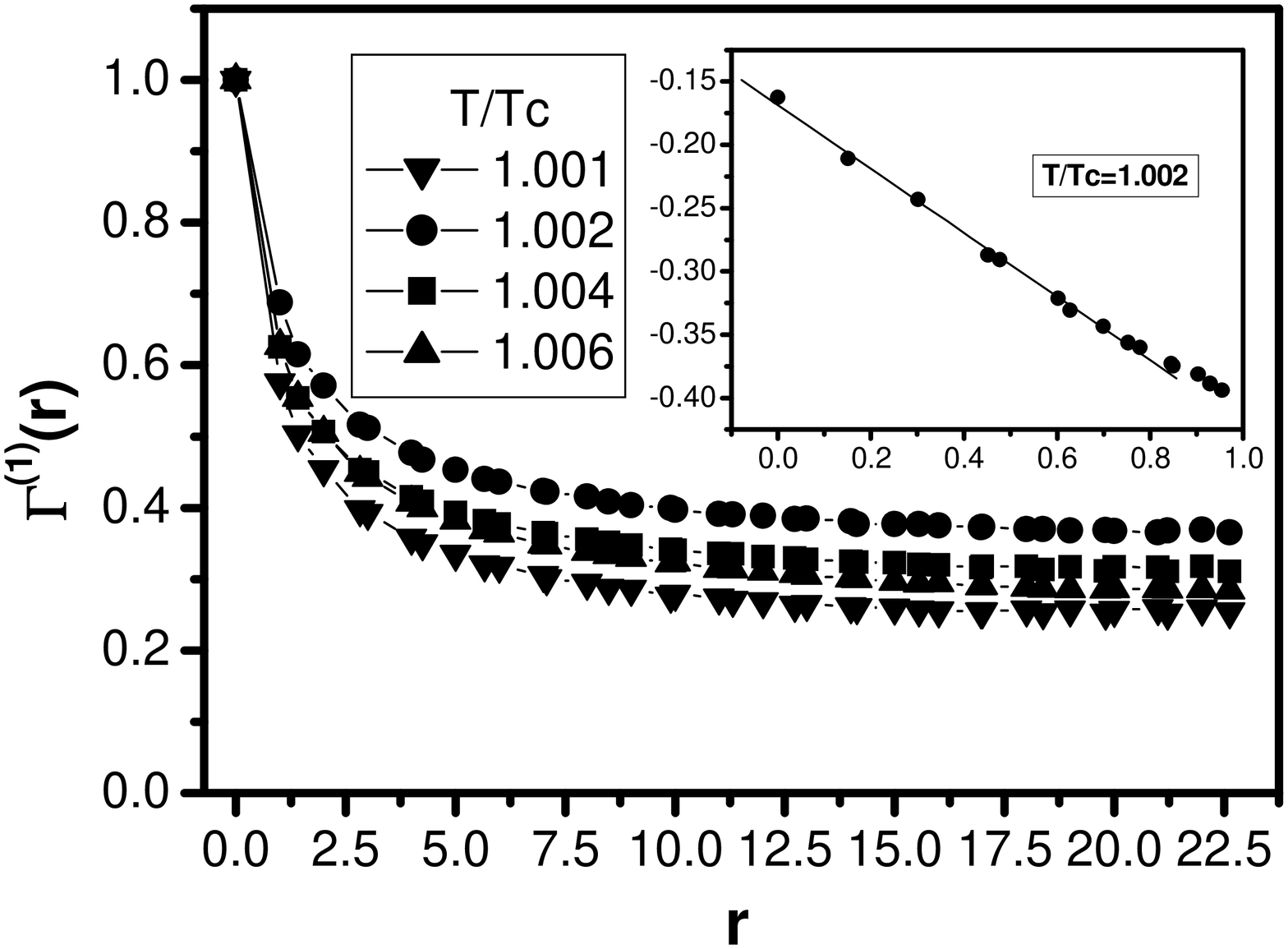}
\end{center}
\caption{\small The correlation function $\Gamma^{(1)}(r)$ versus
$r$. The slowest decay was found for $T/T_{C}=1.002$, in which
case the plot of $\log_{10} [\Gamma^{(1)}(r)]$ versus $\log_{10}
r$ is presented in the inset, leading to the estimate $\eta =
0.\!2521 \pm 0.\!0050$.} \label{fig1}
\end{figure}

In Fig. 1 we present the behavior of $\Gamma^{(1)}(r) $ for
several values of $T/T_C$ (here $T_C$ is the usual Ising critical
temperature at the decoupling point $J_4 = 0$). It is noted that
the slowest decay occurs for $T/T_C = 1.\!002$. The inset in Fig. 1 shows
a log-log plot of $\Gamma^{(1)}(r)$ versus $r$, at the observed
critical point, from which we obtain $\eta = 0.\!2521 \pm
0.\!0050$. Therefore, for the 3-color AT model we found an
evidence of a continuous transition of Ising type for $J_4/J_2 =
0.\!01$. The complete phase diagram for this particular case can
be determined by the present method, as long as the transitions
are continuous. We hope to address to this problem in a future
work.

Recently published results \cite{anjos08,anjos06, anjos07} in other nontrivial situations have established the accuracy of the present approach to obtain critical properties of classical spin models in translationally invariant systems. We thus have presented news exact relations conecting some damage-spreading functions to thermal equilibrium properties for the ferromagnetic N-color Ashkin-Teller model and we hope that such relations will become a useful guide to high-precision numerical investigations.

\end{document}